\documentclass[twocolumn,pra,aps]{revtex4}
\usepackage{graphicx}
\usepackage{amsmath}
\usepackage{amssymb}
\usepackage{centernot}
\usepackage{enumerate}

\begin{document}

\title{Defining what is Quantum:\\Not all what matters for physical phenomena is contained in space-time}

\author{Antoine Suarez}
\address{Center for Quantum Philosophy \\ Ackermannstrasse 25, 8044 Z\"{u}rich, Switzerland\\
suarez@leman.ch, www.quantumphil.org}

\date{May 13, 2019}

\begin{abstract}

It is argued that the three main quantum interpretations, Copenhagen, de Broglie-Bohm, and Many-Worlds, support the \emph{Principle Q (Quantum):} \emph{Not all what matters for physical phenomena is contained in space-time}. This principle underpins Born's rule as well. So \emph{Principle Q} may be the best way to defining Quantum ``from more fundamental principles".
\ \\


\end{abstract}


\maketitle

\section{Introduction}
\begin{figure}[t]
\includegraphics[width=80mm]{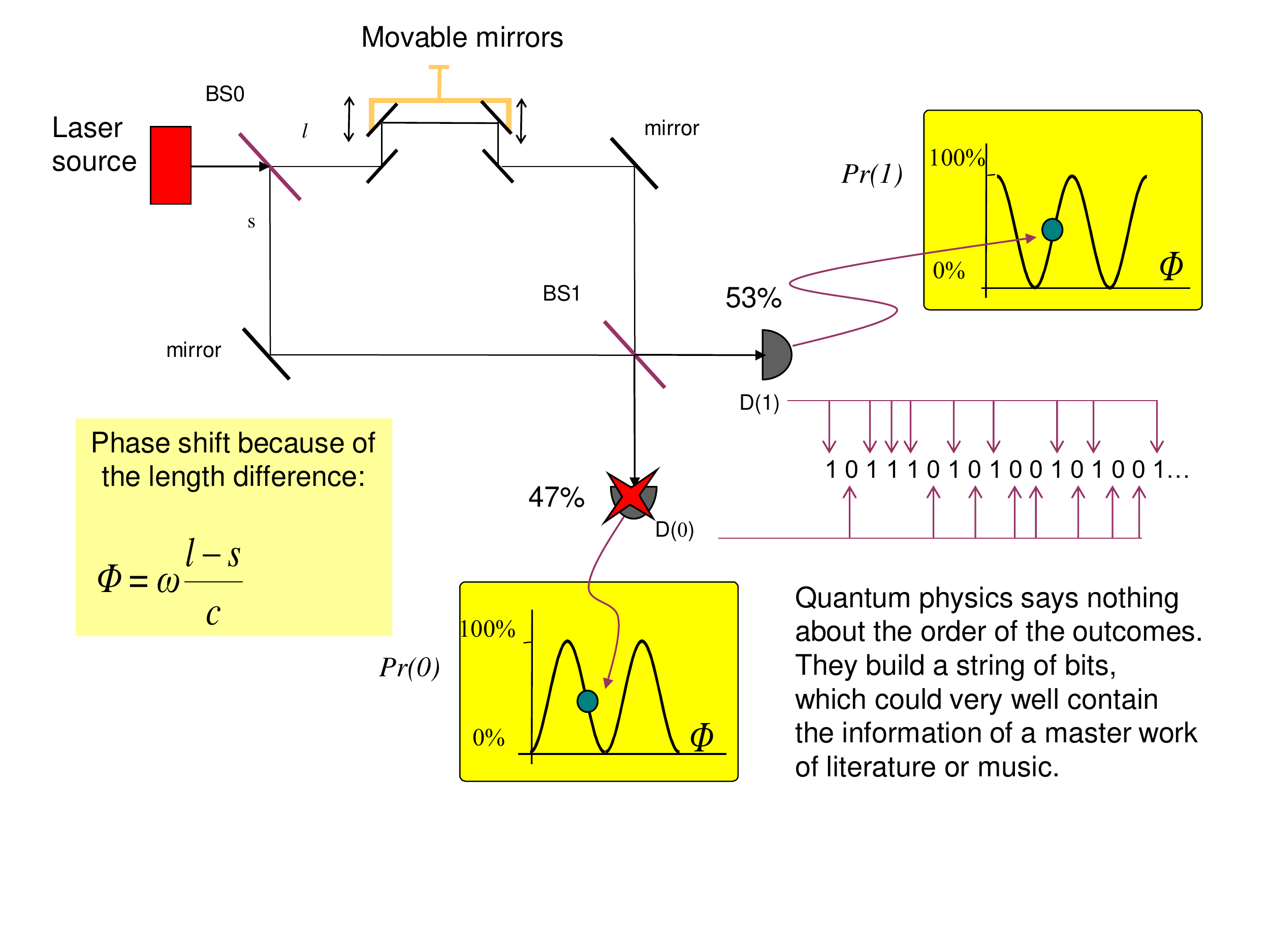}
\caption{Interference experiment: Laser light of frequency $\omega$ emitted by the source enters a Mach-Zehnder interferometer through beam-splitter (half-silvered mirror) BS0 and gets detected after leaving beam-splitter BS1. The light can reach each of the detectors D(1) and D(0) by the paths $l$ and $s$; the path-length $l$ can be changed by the experimenter. For calculating the counting rates of each detector one must take into account information about the two paths leading from the laser source to the detector (\emph{wave behavior}). However, with a single-photon source only one of the two detectors clicks: either D($1$) or D($0$) (\emph{particle behavior}): ``one photon, one count'', or conservation of energy. If $a \;\in\{+1,-1\}$ labels the detection values according to whether D($1$) or D($0$) clicks, the probability of getting $a$ is given by $P(a)=\frac{1}{2}(1+ a \cos \mathit\Phi)$, where $\mathit\Phi=\omega\tau$ is the phase parameter and $\tau=\frac{l-s}{c}$ the optical path.}
\label{MZ}
\end{figure}

The beginning of Quantum is marked by interference experiments with single particles: Double-Slit is paramount. For the sake of our discussion it is convenient to refer to a variant of this experiment using a Mach-Zehnder interferometer as sketched in Fig.\ref{MZ}. The output ports of the second beam-splitter BS1(the output ports of the interferometer) are monitored by corresponding detectors D(0) and D(1): If the outcome (i.e.: which of the two detectors counts) were determined before detection by the path the particle travels, then half of the time D(0) should count and half of the time D(1), and the interference pattern would disappear, what is not the case.

The experiment shows that with sufficiently weak intensity of light, only one of the two detectors clicks: either D(0) or D(1) (\emph{photoelectric effect}). Nevertheless, for calculating the counting rates of each detector one must take into account information about the two paths leading from the laser source to the detector (\emph{interference effect}): The counting rate depends sinusoidally on the optical path-length difference.

The whole ``weirdness" of Quantum is contained in this result:

It is this result and similar ones that led the founding fathers to the insight that in quantum experiments the outcomes are not determined before detection. This insight was strengthened through two main further developments: ``Bell theorem" proving that quantum experiments violate criteria of local causality (Bell's inequalities) \cite{jb}, and ``Kochen and Specker theorem" proving that measurements on quantum systems are contextual (depend on what the experimenter decides to measure) \cite{ks67}.

And it is also worth mentioning that this result defines also the \emph{qubit}. The pioneer quantum technologies, cryptography (BB84-protocol) and computing (Deutsch-algorithm), have been discovered and first technologically implemented using \emph{single qubit interference}.

Nonetheless, although one easily accepts ``Bell's nonlocality" (violation of Bell's inequalities by space-like separated measurements) and ``KS contextuality" as signatures of Quantum, one continues to wonder about what is properly Quantum in single-particle interference, or ask why the qubit is Quantum.

There is some irony in this: Demonstrating Bell's nonlocality involves space-like \emph{multipartite} systems, and hence does not apply to Hilbert spaces with prime dimension; plays a decisive role in 2 qubits systems corresponding to ``four state systems" or Hilbert spaces with $d=4$, but has no significance in single particle spin $3/2$ systems, which also correspond to Hilbert spaces with $d=4$. And contextuality is certainly more general than Bell nonlocality, since it is holds for any system with a number of systems equal or larger than 3 (Hilbert space with $d>2$), however it does not characterize single-particle interference experiments (Fig.\ref{MZ}), and in particular the \emph{qubit}.

On the other hand one may wonder why a so successful theory like quantum physics bears different interpretations whose advocates like to present them as strongly opposite to each other. As it has been pointed out, a good theory should need no interpretation \cite{jf,fp}.

In this letter we will comparatively discuss the three main interpretations Copenhagen, de Broglie-Bohm, and Many-Worlds in the light of the following two principles:

\begin{itemize}
  \item \emph{Principle A (Accessibility):} \emph{All that is in space-time is accessible to observation (except in case of space-like separation).}
  \item \emph{Principle Q (Quantum): Not all what matters for physical phenomena is contained in space-time.}
\end{itemize}

We argue that \emph{Principle A} should be acknowledged by any sound experimental science. On this basis we show that \emph{Principle Q} is somewhat hidden in all the three interpretations: These are not that divergent from each other, but rather different ways of stating the same \emph{Principle Q}. Additionally this \emph{Principle Q} characterises already two-state experiments like that in Fig.\ref{MZ}: The \emph{qubit} is the cornerstone of Quantum, and having it does not require systems with more than 2 states (Hilbert space $d>2$). So we conclude proposing to found Quantum on \emph{Principle Q}.

In searching for ``more fundamental principles" behind quantum formalism Max Born provides a useful compass: ``[it] is a philosophical question for which physical arguments alone are not decisive."\cite{mb26}

\begin{figure}
\includegraphics[width=50mm]{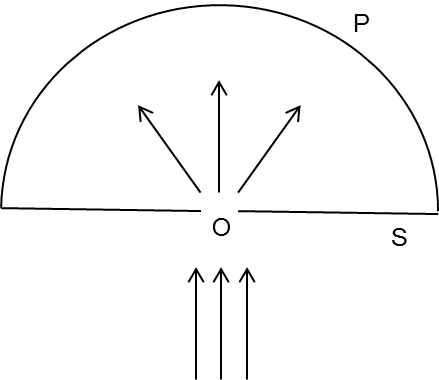}
\caption{Einstein's gedanken-experiment: Let S be a diaphragm provided with a small opening O, and P a hemispherical photographic film of large radius. Electrons impinge on S in the direction of the arrows. ``There are de Broglie waves, which impinge approximately normally on S and are diffracted at O. Behind S there are spherical waves, which reach the screen P and whose intensity at P is responsible [massgebend] for what happens at P.'' (See \cite{bv} p. 486: in the text both 'P' and 'S' are confusedly referred to as ``screen'' (\'{e}cran)).}
\label{Sol}
\end{figure}

\section{Copenhagen} \label{copenhagen}
Single particle interference led to the Copenhagen (``orthodox") interpretation, which considers crucial the moment of detection and postulates a so called ``wavefunction collapse''. This rather cryptical name conflates two different assumptions:

a) The decision of the outcome (which of the two detectors counts) happens at the moment of detection.

b) At detection the outcome becomes ``irreversibly recorded" and can be observed.

The different pictures discussed in the following regard assumption a). Interpretations regarding assumption b) will be discussed in a separated paper and will allow us to complete the definition of Quantum.


Assumption a) implies ``nonlocal" coordination of detection events, or more accurately of decisions at detectors between ``to count" or ``not to count".

As it is well known, Einstein argued against this ``nonlocality at detection" already in 1927, at the 5$^{th}$ Solvay conference in Bruxelles \cite{bv}. He did this on the basis of the famous gedanken-experiment in Fig.\ref{Sol} concluding (erroneously) to a conflict with relativity:

\hangindent=0.36cm
\hangafter=0
\noindent
{\footnotesize ``But the interpretation [II], according to which $|\psi|^2$ expresses the probability that \emph{this} particle is found at a given point, assumes an entirely peculiar mechanism of action at a distance, which prevents the wave continuously distributed in space from producing an action in two places on the screen. In my opinion, one can remove this objection only in the following way, that one does not describe the process solely by the Schr\"{o}dinger wave, but that at the same time one localises the particle during the propagation. I think that Mr de Broglie is right to search in this direction. If one works solely with the Schr\"{o}dinger waves, interpretation II of $|\psi|^2$ implies to my mind a contradiction with the postulate of relativity.''}\cite{bv}
\\

Here Einstein is referring to Max Born's interpretation in his seminal paper in 1926 \cite{mb26}, and the gedanken-experiment Einstein proposes (Fig.\ref{Sol}) is a simplified version of the scattering of an electron by an atom Born discusses in his paper. Einstein sharply perceived that Born's interpretation is linked to nonlocal coordination of detectors. In fact it was the impossibility of explaining this coordination by ``causal evolution" what motivated Born to his celebrated statistical interpretation, as we discuss later in Section \ref{born}.

\begin{figure}
\includegraphics[width=87mm]{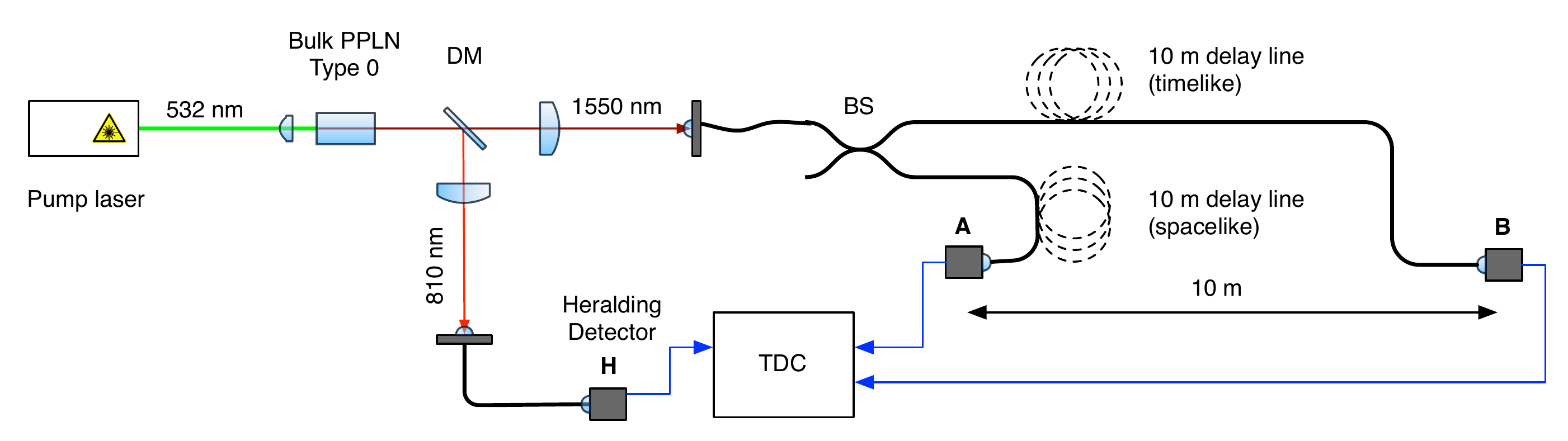}
\caption{Experimental setup to demonstrate nonlocal coordination between decisions at detectors (in accord with Einstein's gedanken-experiment in Fig.\ref{Sol}): photon pairs are generated by Spontaneous Parametric Down Conversion at the wavelengths of 1550nm and 810nm. These pairs are split by a dichroic mirror (DM): the 810nm photon is sent to detector D, used to herald the presence of the 1550nm photon, which follows to the beam splitter (BS).\cite{Guerreiro12}}
\label{Su}
\end{figure}

Astonishingly Einstein's gedanken-experiment in 1927 has been first realized using today’s techniques in 2012 \cite{Guerreiro12}. In this experiment (Fig.\ref{Su}) single photons impinge into a beam-spitter BS and thereafter get detected. The two detectors A and B monitoring the output ports of BS are located so that the decision ``to count" or ``not to count" at A is space-like separated and therefore \emph{locally} independent from the decision ``to count" or ``not to count" at B. The experiment tests and rules out the assumption that this correlation can be explained by some sort of local coordination through signals with $v\leq c$: Even if the decision at A is \emph{locally independent} of the decision at B, both decisions yield correlated results. On the other hand this nonlocal coordination cannot be used to signal faster than light from one detector to the other. The experiment also highlights something Einstein did not mention: Nonlocality is necessary to preserve such a fundamental principle as \emph{energy conservation} \cite{Guerreiro12}.

Einstein understood that the quantum mechanical description of single particle experiments involves nonlocality, but was not ready to swallow this.

\section{de Broglie-Bohm}\label{dbb}
Another important player at the 5th Solvay Conference was Louis de Broglie. He presented an interpretation different from the Copenhagen one, the so called ``pilot-wave" picture: In the experiment of Fig.\ref{MZ} the particle always follows a well determined path from the source to a detector, but there is a ``pilot wave", an \emph{undetectable} mathematical entity, that guides the particle to one or other of the detectors taking account of the optical path-length difference to producing the characteristic interference fringes predicted by Quantum Mechanics.

Unfortunately Einstein misinterpreted de Broglie's ``pilot wave" as a local alternative to the nonlocal ``Schr\"{o}dinger waves" (the quantum mechanical ``wavefunction"), as the quotation in the preceding Section \ref{copenhagen} reveals. The kind of ``wave" Einstein had in mind was sort of ``ghost-wave" propagating in ordinary space-time by the alternative path to the localised path the material particle takes: it carries neither energy nor momentum, and travels at the same speed of the particle. Thereby one can assume that the outcome becomes determined when the ``ghost-wave" joins the particle at beam-splitter BS1, and explain interference avoiding nonlocal coordination of decisions at detectors. Nonetheless one pays the price of admitting entities propagating in space-time that are inaccessible to observation even in principle. In other words, to save relativistic local causality Einstein was unconsciously discarding the \emph{Principle A}, which is in fact the fundamental principle of any experimental science.

Eight years later (1935) Einstein proposed a new thought experiment with two particles in the famous EPR paper \cite{epr}. It looks like if Einstein had smelled that his ``ghost-waves" do not allow to escape nonlocality in entanglement experiments. It may be also that he was not ready to get rid of \emph{Principle A} after all. As a matter of fact he never endorsed definitely the ``ghost-wave" model and abandoned his 1927 argument in favor of the more complicated EPR one, arguing that Quantum Mechanics cannot be considered complete. But then the question arises: Which kind of variables was Einstein searching for to complete Quantum Mechanics? \cite{epr}

In whichever way one looks at it one can't help concluding he was looking for variables which are observable in principle even if they couldn't be detected by available experimental techniques. In this sense it has been claimed Einstein was supporting an ``epistemic" view that considers the quantum probabilities as ``lack of knowledge" \cite{hasp}. This ``epistemic view" bears a problem: If the variables are in principle accessible to observation, the experimenter could in principle know in advance which path the particle will take, and thwart the appearance of interferences by changing the length of the other path. Thus, to fit interference experiments ``epistemicism" has to assume that nature coaxes the experimenter to adapt his/her choices to the properties the particle carries. At the end of the day Einstein was denying the freedom of the experimenter, that is, endorsing \emph{Superdeterminism}.

Notice that Copenhagen can also be considered an ``epistemic" description, in the sense that the ``wavefunction" describes only the possible knowledge we can have about the ``system", but does not contain information about the outcome of a single quantum event. In Bohr's wording: ``There is no quantum world. There is only an abstract quantum physical description. It is wrong to think that the task of physics is to find out how nature is. Physics concerns what we can say about nature." However there is a big difference: ``Bohr's epistemicism" acknowledges that the information lacking in the ``wavefunction" is inaccessible to observation in principle and, therefore, is not contained in space-time. If by ``world" one refers to ``what is contained in space-time", then ``there is no quantum world", but there is a mathematical quantum reality existing in a mental realm we cannot access with our senses. The ``epistemic" description either endorses \emph{Superdeterminism} or \emph{Principle Q}.

Seventeen years after the EPR-paper David Bohm published (1952) an article applying de Broglie’s picture to describe the EPR gedanken-experiment \cite{db}. In this context it became clear that the ``empty-wave" has nothing to do with a ``ghost-wave" propagating locally within the ordinary 3-space, but is rather a mathematical entity defined in the so called ``3N-space or configuration space"(\cite{jb} p. 128). Additionally, the wave acts like a quantum potential involving action at a distance with \emph{infinite} velocity to produce space-like separated correlated detections at the two sides of the setup. Because de Broglie-Bohm postulates a ``preferred frame" defining a ``before-after" relationship between two space-like correlated events it conveys the idea of \emph{nonlocal} action at a distance: one of the events is the cause and the other the effect. However experimenters cannot use this ``effect" to communicate faster than light. This amounts to say that the ``before-after" relationship is a quality inaccessible to real measurement, and therefore not contained in the observable world, that is, space-time. In other words, the ``wave'' guides the ``material particle" from outside the ordinary space-time, i.e.: nonlocally.

The idea of causal action at a distance in the Bohmian picture inspired John Bell to find a criterion for deciding between Bohmian quantum nonlocal causality and Einstein’s relativistic local one by means of entanglement experiments: The meanwhile famous Bell inequalities \cite{jb}. Since 1982 a number of well-known experiments have ruled out Einstein’s locality, upholding the quantum mechanical predictions with increasing accurateness. Additionally, Bohmian picture inspired the ``before-before" experiment \cite{szgs}, which ruled out the nonlocal causal view that ``one event happens first and the other after", and led to the insight that ``the correlations come from outside space-time" \cite{ng12}.

Ironically, because of the ``prejudice" of temporal causality de Broglie-Bohm could better than Copenhagen bring into focus that quantum correlations cannot be explained by signals bounded by the velocity of light, and triggered ``nonlocality research".

One may be tempted to think that de Broglie-Bohm is ``less weird" than Copenhagen since at least the former assumes well localized material particles and thereby an ontological substrate. This would mean to overlook Louis de Broglie's big discovery that material particles behave like waves: Particles are localized material entities only apparently, actually they consist in immaterial waves. It is the very fundamental idea of the ``wave-packet". Shaping a ``wave-packet" means weaving ``local particles" with ``nonlocal threads" (plane-waves) using the tool of ``Fourier transforms". The idea that ``local" and ``nonlocal" are inseparably united in the quantum phenomena is undoubtedly the main characteristic of the ``particle-pilot wave" model. Nonetheless the idea was already present in de Broglie's ``wave-packet" description for particles, and indeed in a way which was fully acknowledged and worked out by the Copenhagen founding fathers:

\begin{itemize}
  \item Describing interference in experiments like that of Fig.\ref{MZ} requires the notion of ``wave-packet" characterized by a coherence-time and coherence-length, and yields the sinusoidal dependence on phases characteristic of quantum correlations \cite{as10a}. Depending on the optical path-length difference the same ``wave-packet" behaves like if it were a ``wave" producing phase-dependent counting rates, or like if it were a ``particle" producing phase-independent counting rates.
  \item The impossibility of sharply measuring both a variable and its Fourier transformed is the root of Heisenberg's uncertainty principle.
  \item ``Single particle interference" implies that each jointly outcome of the corresponding detectors should be considered a single measurement result coming from outside space-time: Different phases define in general different bases and different measurements incompatible with each other.
  \item De Broglie's idea of the electron as wave-packet plays a crucial role in Max Born's description of collision processes, where he introduces his rule for calculating probabilities of outcomes in quantum experiments \cite{mb26}. Born's paper shows well that de Broglie's view supports Copenhagen: Quantum ``particles" like an electron and an atom are abstract mathematical descriptions that combine to build a compound wave-function describing the scattering of an electron by an atom. So particles exist in a mental realm till they become materialised in one of different possible experimental outcomes (we comment on this more extensively in Section \ref{born}).
\end{itemize}

The important lesson of the whole story is that the ``de Broglie-Bohm" interpretation is itself essentially nonlocal, although it can be misunderstood as local in single-particle experiments, as Einstein did. Copenhagen and de Broglie-Bohm are two different ways of stating \emph{Principle Q}, rather than two interpretations irreconcilable with each other.

\section{Many-Worlds}
The strongest reaction against the ``quantum collapse" came from the so called Many-Worlds interpretation (MWI).

This interpretation goes back to the relative state formulation of Quantum Mechanics proposed by Hugh Everett 1957 \cite{he}. ``The fundamental idea of the MWI [...] is that there are myriads of worlds in the Universe in addition to the world we are aware of. In particular, every time a quantum experiment with different possible outcomes is performed, all outcomes are obtained, each in a different world, even if we are only aware of the world with the outcome we have seen."\cite{lev}.

The parallel worlds and observers resulting at each quantum experiment are in principle ``experimentally" inaccessible to each other.

Many-Worlds has been formulated in various ways. David Deutsch uses the ``Multiverse" formulation to explain the notion of a ``quantum computer" \cite{deutsch}. The meaning of Many-Worlds is particularly well brought to light in the formulation ``Parallel lives" by Gilles Brassard and Paul Raymond-Robichaud \cite{Brassard}. According to this version: ``When Alice pushes a button on her box, she splits in two, together with her box. One Alice A sees the red light flash on her box, whereas the other A* sees the green light flash. Both Alices, A and A*, are equally real. However, they are now living parallel lives: they will never be able to see each other or interact with each other. In fact, neither Alice is aware of the existence of the other, unless they infer it by pure thought as the only reasonable explanation for what they will experience when they test their boxes." \cite{BRR1}

Variants and extensions of Many-Worlds are properly characterized by Frauchiger-Renner's statement: ``Their common feature is that they do not postulate a physical mechanism that singles out one particular measurement outcome, although observers have the perception of single outcomes."\cite{fr} This characterization points also to what may be a main inconsistency of Many-Worlds:

Indeed, according to Leibniz's principle, ``if there is no possible perceptible difference between two objects, then these objects are the same, not superficially, but fundamentally" \cite{BRR2}. As far as one keeps to this principle, if Alice in our world can never be able to see the other Alice*, then one should conclude that Alice* has no physical reality at all: Things that cannot in principle be perceived by the senses do not exist \emph{within space-time}. And if the existence of Alice* can be inferred by reasoning but cannot in principle be perceived by the senses, this means that Alice* exists outside space-time.

But one could object: Wouldn't the latter statement imply that events that happened in the distant past, about which we just read in history books, but which we cannot (and could not) perceive with our senses, do not exist?

We access such events through observations we perform today: archaeological vestiges or writings documenting them (see British museum).
So these events happened in our space-time. A past event that is in principle inaccessible to any observation we can perform does not belong to the physical reality we can describe.

The Alice* in the ``parallel-lives" interpretation is something we cannot access through any observation in our world but only through reasoning. Accordingly, Alice* lives outside our space-time, and is inaccessible to our senses. This amounts to say that Alice* is a mental entity outside my observable reality.

``Parallel lives" makes clear that ``Copenhagen" and ``Many-Worlds" are basically equivalent: Both interpretations imply that the ``physical reality" we live in is more than what we can access with our senses. We discuss this further in the coming Section \ref{apw}.

\section{Ernst Specker's ``Infuturabilien" and All-Possible-Worlds}\label{apw}
Consider a conventional Bell experiment: On one side of the setup Alice measures by switching her apparatus either on position $a_{1}$ or $a_{2}$, and gets outcome either 1 or 0 in each measurement; on the other side far away, Bob measures by switching his apparatus either on position $b_{1}$ or $b_{2}$, and gets outcome either 1 or 0 in each measurement. Suppose a measurement with joint choice [$a_{1}$,$b_{2}$] yields the coincidence outcome [1,0]. According to Many-Worlds the other possible outcomes [1,1], [0,1], [0,0] are realized in parallel worlds and observed by corresponding clones of Alice and Bob.

However, Many-Worlds to be consistent should also consider all the possible choices Alice and Bob can do, and assume that the other three possible joint choices [$a_{1}$,$b_{1}$], [$a_{2}$,$b_{1}$], [$a_{2}$,$b_{2}$] are realized in parallel worlds as well. Paraphrasing Nicolas Gisin: In Many-Worlds the experimenter should not be merely a passive observer, but play an active role \cite{ng17}. Accordingly, in each experiment the world would split in 16 parallel worlds.

But then one could as well think of a ``razored" version keeping only the parallel worlds corresponding to the different choices experimenters can do, and renounce to the splitting corresponding to all the possible outcomes that could happen in each choice.

This completed version of Many-Worlds leads straightforwardly to our proposal of ``All-Possible-Worlds" \cite{as17}: In the context of Ernst Specker's ``infuturabilien" parable \cite{Specker60} one can consider that an ``omniscient mind" assigns a well-defined outcome for each of the four possible choices Alice and Bob can make. For different rounds of a same experiment (say [$a_{1}$,$b_{2}$]) assignments are done so that after many rounds the experimenters will observe joint outcomes distributed according to the quantum mechanical ``Born's rule", and hence violating Bell's inequalities.

``All-Possible-Worlds" illustrates well what quantum contextuality means: The omniscient mind assigns results to each possible choice and each possible round, however the assignment is not done for each single choice of Alice and each single choice of Bob \emph{separately}, but \emph{to each possible pair of choices jointly}. Consequently for each round the outcome assignment for Alice's choice (say $a_{1}$) depends on whether Bob chooses $b_{1}$ (to perform experiment [$a_{1}$,$b_{1}$]) or $b_{2}$ (to perform experiment [$a_{1}$,$b_{2}$]).

This also means that the outcome (the ``collapse", understood in the sense of Assumption a) in Section \ref{copenhagen}) is actually the result of two decisions, that of the omniscient mind (when this assigns outcomes to all possible experiments) and that of the experimenter (when this decides to perform a determined experiment). Detection (the measurement) does not ``single out" the outcome, it rather makes it visible.

Many-Worlds has the enormous merit of having brought to light an important idea hidden in the Quantum: The physical reality consists in all possible experiments (choices) humans of all times can perform, the number of which is breathtaking huge but \emph{finite}, as explained in \cite{as17}. These choices and the corresponding outcomes are all present in God’s mind \cite{aa} but I am free to choose the world I want to live in.

As David Deutsch states: ``we find ourselves unavoidably playing a role at the deepest level of the structure of physical reality."\cite{deutsch} And paraphrasing Scott Aaronson: The mind of God works like a ``tensor factor in Hilbert space" to assigning outcomes to all possible rounds of all possible experiments. Observed phenomena reveal what the state of this mind is with relation to the particular choices we freely make \cite{sa17}.

Some remarkable implications of ``All-Possible -Worlds" have been discussed in \cite{as17}.

\section{Free-Will as axiom of Quantum}\label{kochen}
The conclusions in the preceding Sections can be strengthened on the basis of Simon Kochen's reconstruction of quantum mechanics \cite{sk15, sk17}. This reconstruction illustrates well that any interpretation assuming free-will acknowledges \emph{Principle Q.}

The corner stone of Kochen's reconstruction is found in the following quotation referring to EPR experiments:

``How can correlations between spin components of two particles subsist when these spin components do not have values?
To understand how this can happen it is necessary to distinguish between events that have
already happened and future contingent events. Thus, for instance, if $a \vee b$  is currently true, then
either \emph{a} is true or \emph{b} is true. When future events are considered, this no longer the case: if $a \vee b$
is certain to happen, it is not the case that \emph{a} is certain to happen or \emph{b} is certain to take place."\cite{sk17}

As a famous example of this logical feature Kochen refers to Aristotle's sea battle: ``A sea-fight
must either take place to-morrow or not, but it is not necessary that it should take place to-morrow,
neither is it necessary that it should not take place, yet it is necessary that it either should or
should not take place tomorrow. Since propositions correspond with facts, it is evident that when in
future events there is a real alternative, and a potentiality in contrary directions, the corresponding
affirmation and denial have the same character." \cite{sk17}

Here one implicitly introduces the following

\emph{Definition:}
An event \emph{a} is called future contingent with relation to a future time T, if to establish whether event \emph{a} takes place or not one has necessarily to await time T.

It is interesting to compare Aristotle's example of the ``sea battle" with the statement: ``The Sun will raise tomorrow at time T at point P of the horizon".

The ``sea battle" is a future contingent event with relation to ``tomorrow".
By contrast, according to classical physics the raising of the Sun is not a future contingent event because it is assumed we can know today about it with certainty.

According to the \emph{Definition} above it is impossible even in principle to predict with certainty whether event \emph{a} will take place or not on the basis of the currently observable data, that is on the basis of information stored within space-time.

This is the same as assuming that on the basis of \emph{classical} physics we can establish neither that ``\emph{a} will happen at time T" nor that ``\emph{a} will not happen at time T". In other words the \emph{Definition} of future contingents is equivalent to assuming that whether event \emph{a} happens or not at time T depends on information coming from outside space-time at time T. And this is nothing other than assuming \emph{Principle Q}.

The classical view implies actually that no event is future contingent: Classically, any event results \emph{necessarily} through causal evolution.

In case of Aristotle's sea battle \emph{Principle Q} amounts to postulate human free will, i.e.: steering of outcomes in human brains happening from outside space-time; this might also be the reason why Thomas Aquinas postulates that ``the universe would not be perfect without randomness" \cite{vs}. Quantum experiments (single-particle Mach-Zehnder interference, single-particle spin 1 Stern-Gerlach, EPR-Bell experiments) allow us to go beyond: We postulate free will on the part of the experimenter and observe correlations between space-like separated detection events; thereby we experimentally demonstrate information coming from outside space-time in devices other than human brains. This is the very meaning of ``the Free-Will Theorem" \cite{sk17,ck}.

In this very sense Nicolas Gisin emphasizes that Free-Will is an axiom of physics. Interestingly Gisin relies on the French philosopher Jules Lequyer \cite{ng15}, who was strongly motivated by harmonizing divine omniscience and future contingent events, the ``Infuturabilien" question which later inspired Erns Specker as well (Section \ref{apw}).

It is noteworthy that Kochen's reconstruction of quantum mechanics is based on single particle Mach-Zehnder interference experiments and introduces the sinusoidal dependence on phases \cite{sk15}. Even if no physical motivation is given for such a dependence, Kochen's reconstruction makes it plain that any attempt to get Quantum from principles assumes (openly or hiddenly) nonlocal ingredients and the construction of ``wave-packets" by means of ``Fourier transforms".

In summary Kochen's ``logical" reconstruction of quantum physics rests on \emph{Principle Q}. So one can't help admiring how much quantum physics was already contained in Aristotle's logical description about the ``sea battle". Quantum is intrinsically related to those assumptions allowing us to shape rationally our ordinary daily life.

\section{Meaning and authorship come from outside space-time}\label{meaning}
All the different interpretations of Quantum either hiddenly deny the freedom of the experimenter (and therefore lead to Superdeterminism and are not proper interpretations of Quantum), or acknowledge \emph{Principles A and Q.}

Consider now the outcomes my brain produces while I am typewriting this article. One could compare the brain to the quantum interferometer in Fig.\ref{MZ}: Depending of the physiological parameters of my brain an experimenter may be capable of predicting how many percent of each character of the alphabet (`a', `b', `c', …) there will be in the final text (similarly as he/she predicts e.g.: 57\% of bit `1' and 43\% bit `0' in the experiment of Fig.\ref{MZ}). By contrast there is no science (and there never will be one) capable of predicting the order of the characters in the final text (the particular bit-string 1,1,0,0,0,1,1,... resulting after many rounds in the same experiment), and therefore the message I want convey. I mentally steer my brain's outcomes, that is, the sequences of bits they consist in, from outside space-time. This order, which is inaccessible in principle to observation, is what communication is all about and makes our life meaningful.

Postulating that the order the outcomes appear is beyond what physics can predict amounts to acknowledge that authorship responsible for this order is outside space-time. We assume such an authorship when we take for granted \emph{conservation of personal identity} in daily life. This conservation is so fundamental that without it any legal and social order would  break down: We could neither claim for rights nor keep bank accounts. We assuming it all the time while publishing scientific articles and commenting each other's papers. ``Identity" as my agreement to identify with the person named on my passport, driver’s license, bank and google accounts \emph{each and every morning}, is more than a world-line; reducing my life to time-flow mean condemning myself to be none.

If there is no theory capable of predicting each single choice a ``human experimenter" decides to make or the outcome the omniscient mind assigns to each single round of the experiment we choose to perform, then there is no theory of everything. We can only find theories allowing us to predict statistical distribution of outcomes. But as Chris Fuchs wisely states: ``Finding a theory of ``merely" one aspect of everything is hardly something to be ashamed of: It is the loftiest achievement physics can have in a living, breathing nonreductionist world." \cite{caf10}

\section{How did Max Born get to ``Born's~rule"?}\label{born}

In his 1926 paper Max Born introduced a rule, which became an essential ingredient of quantum physics: The amplitudes (``Ausbeutefunktionen") appearing in the Schr\"{o}dinger wave-equation are linked to probabilities of different possible alternative outcomes \cite{mb26}.

Born's rule was not simply a good guess coming from nowhere, or a postulate without further justification. It emerged from the struggle to account for basic physical conditions in the problem of single electrons being scattered by atoms \cite{mb26}, an experiment which as said (Section \ref{copenhagen}) Einstein simplified to the gedanken-experiment in Fig.\ref{Sol} to argue that Born's interpretation implies nonlocal coordination at detection.

For the sake of simplicity we reproduce the essential of Born's argument in the context of the experiment in Fig.\ref{MZ} using current standard notation:

The quantum mechanical description in terms of a wavefunction leads to an expression of the form:
\begin{equation}\label{psi}
| \psi \rangle=c_{0} |0\rangle + c_{1}|1\rangle
\end{equation}

where $|0\rangle$ and $|1\rangle$ denote the outcome 'D(0) counts', respectively 'D(1) counts'; and $c_{0}, c_{1}$ are complex numbers resulting from summing amplitudes over paths from the source to detector D(0), respectively D(1).

According to the ``corpuscular" picture one gets only one outcome: Either D(1) counts or D(0) counts. However in the right hand side of Equation (\ref{psi}) appear together the two possible outcomes: D(1) counts and D(0) counts. So Equation (\ref{psi}) does not describe a causal evolution from the source to a single detector: ``From the standpoint of our quantum mechanics there is no quantity that in any individual case causally fixes the consequence of the collision."\cite{mb26} This amounts to say that the conditions ensuring that only one of the two detectors counts lay beyond space-time and are inaccessible in principle. They may derive from ``a preestablished harmony" (``eine pr\"{a}stabilierte Harmonie") \cite{mb26}.

Born's is facing the following dilemma: On the one hand to behave rationally in the world we need receipts allowing us to predict things to some extent. On the other hand the fundamental fact that the outcome's assignment is inaccessible in principle because it is not contained in space-time and does not emerge through causal evolution entails that there is no receipt allowing us to predict each single outcome. The only possible interpretation is that Equation (\ref{psi}) does not allow us to predict with certainty but only to make a guess. Probability is a measure of how accurate such a guess may be. So we have to derive these probabilities from the quantities $c_{0}, c_{1}$ in Equation (\ref{psi}), Born concludes.

So, Born's claim that quantum physics ``only specifies probabilities, and not definite outcomes" \cite{ma} is derived from ``more fundamental principles", mainly the need to account for correlated decisions at detection ``in absence of conditions ensuring a causal evolution". In stating that ``the only possible explanation" of the ``wave function" is probabilistic Max Born was in fact assuming \emph{Principle Q}: Quantum phenomena cannot be explained exclusively by local causality but require ``nonlocality at detection", as Einstein well understood (Section \ref{copenhagen}).

Stating that quantum physics is ``intrinsically probabilistic" \cite{jf} amounts to state that Quantum refers to a realm which is not fully contained in space-time and precludes explanation by causal evolution.

Born's interpretation is amazingly close to the picture ``All-Possible-Worlds" described in Section\ref{apw}:

\begin{itemize}
  \item The ``preestablished harmony" corresponds to the assignments the omniscient mind makes for all possible experimental choices humans of all times can freely make (a \emph{finite} number as explained in \cite{as17}).
  \item ``Giving up determinism" corresponds to the assumption that humans can freely chose which of the possible worlds they want to live in.
  \item ``Absence of conditions ensuring a causal evolution" means that the state of the omniscient mind is outside space-time, i.e.: contains correlated events that are not predetermined by the past light-cone.
\end{itemize}

From the perspective of the omniscient mind, Born's rule probabilities mean the frequency of occurrences of an outcome in the sequence of assignments this mind does, which is the actual observed frequency in experiment. In this sense the rule uses \emph{``frequentist probabilities"} and can be considered \emph{objective} (as \emph{\textsf{BornF}} in \cite{fr17}).

From the perspective of the human experimenter, Born's rule quantifies a subjective belief about the outcome of a future measurement under conditions of irreducible uncertainty \cite{fp}. This interpretation (characteristic of QBism \cite{caf,caf17}) reflects the incapacity of principle to access the result of a measurement before performing it, and uses \emph{``Bayesian probabilities"}. In this sense Born's rule can be considered \emph{subjective} (as \emph{\textsf{BornB}} in \cite{fr17}).

For all practical purposes both interpretations are equivalent, since the assignments in the omniscient mind are done in such a way that the factual frequencies occurring in experiment are well fitted by the Bayesian probabilities. This ``preestablished harmony" ensures that ``[we] can make rational decisions in the face of uncertainty"\cite{fp}, and behave rationally in the world we live in. So the \emph{objective} view and the \emph{subjective} become unified in ``All-Possible-Worlds".

In the same line of thinking one could also say that Max Born was introducing probabilities to face ``lack of knowledge". Nonetheless not ``knowledge" we would be able to get hold of with better apparatuses, but one that is inaccessible to any apparatus we can built. By contrast Boltzmann thermodynamics  probabilities appear as a convenient tool to calculate a situation where it is very difficult but \emph{not impossible} to take account of each single case. Interestingly at the end of his paper Born feels the necessity to stress the difference between his ``statistical interpretation" and the ``thermodynamical-statistical" one.

So, stating that quantum physics is ``epistemic" \cite{hasp} amounts to acknowledge \emph{Principle Q} once again.

Furthermore Born's reasoning is based on de Broglie's idea that a particle can be considered a wave packet obtained by summing up plane waves of different frequencies according to a Fourier transform relation (Section \ref{dbb}). The very introduction of ``wave packets" amounts to assume (paraphrasing Wheeler) \emph{(local) 'it' from (nonlocal) 'qubit'} as primitive for Quantum. This means that the ``quantum algebra" (leading to violation of Bell inequalities, ``Tsirelson bound" and contextuality) emerges in the context of experiments like that of Fig.\ref{MZ} \cite{as10a}. In particular the sum of the probabilities for the counting rates of D(1) and D(0) has to be 1, and hence in Equation (\ref{psi}) it holds that: $ |c_{0}|^2 + |c_{1}|^2=1$ (\emph{unitary transformations}); and in an experiment with the detectors watching BS0 instead of BS1, the algebra must lead to the classical probabilities. Such algebraic properties of the wavefunction formalism led Born to the famous footnote correction (in \cite{mb26}, p. 865) that the probabilities are given by $|c_{i}|^2$, $i \;\in\{0,1\}$, and not by $c_{i}$.

From this perspective ``nonlocal coordination of decisions at detectors" bears also ``contextuality" (characterizing Hilbert spaces with $d \geq 3$) and ``Bell's nonlocality" (characterizing Hilbert spaces with $d \geq 4$, $d$ non prime), and so rules the whole quantum realm (Hilbert space with $d \geq 2$).

In summary: Max Born did not introduced his rule as a ``probabilistic axiom". He derived it from more fundamental principles ``without relying on an \emph{a priori} notion of probabilities" \cite{fr17}, mainly from ``absence of conditions ensuring causal evolution", irreducible uncertainty, and the Hilbert space properties embedded in the description of particles as``wave-packets" (\emph{it from qubit}).

One can easily see that the derivation of Born's rule from three ``operational postulates or primitives" in \cite{mgm} contains much of Born's reasoning we have referred to. This becomes clear if one does not forget to introduce detectors into the definition of ``measurement": Measurements without detectors have no outcomes. Mateus Ara\'{u}jo remarks that \cite{mgm} leaves hanging the question: ``Why does the Born rule only specifies probabilities and not definite outcomes?" \cite{ma}. In fact the question could be answered because ``absence of conditions assuring a causal evolution" or ``preestablished harmony" is implicitly assumed in the postulate ``\emph{states}".

It is also noteworthy that in \cite{fr17} the authors claim to derive Born's rule from ``two alternative non-probabilistic physical postulates" together with ``certain natural assumptions concerning the agent's reasoning" that ``do not depend on quantum mechanics". As we have seen, this is actually what Max Born did: Assumption \textsf{\emph{``Symmetry"}} in \cite{fr17} states that ``the agent’s belief about a sequence of
outcomes obtained by repeating the same prepare-and-measure experiment does not depend on how the sequence is ordered"; this amounts to assume that the order of the sequence is beyond what physics can predict and therefore ``absence of conditions ensuring causal evolution" (Section \ref{meaning}). Assumption \textsf{\emph{``Overlap"}} implicitly involves nonlocal coordination at detection. And Assumption \textsf{\emph{``Repeat"}} implicitly contains the Hilbert space algebra to calculate a path's amplitude.

\section {Conclusion}

As early as 1927 Einstein was perfectly aware that single-particle nonlocality, later demonstrated in the experiment of \cite{Guerreiro12}, is the source where Quantum emerges from. His denial of this feature ignited a debate that crystalised into three main interpretations of quantum theory: Copenhagen, de Broglie-Bohm, and Many-Worlds. In this letter we have shown that these interpretations are different ways of stating the same from different perspectives: Not all what matters for physical phenomena is contained in space-time (\emph{Principle Q}).

The three interpretations highlight different relevant aspects, which can be unified into the ``All-Possible-Worlds" picture: The quantum realm is a huge collection of All-Conceivable-Histories. Paraphrasing John A. Wheeler: The entirety of quantum phenomena, rather than being built on particles or fields of force or multidimensional geometry, is built upon billions upon billions of elementary human decisions (\cite{caf17} p.32, note 35). Without ``human free choices", no physical reality! Personal identity and free will are axioms of science included in \emph{Principle Q}, natural assumptions about rational reasoning and behavior.

\emph{Principle Q} underpins Max Born's introduction of probabilities into quantum physics. The founding fathers undoubtedly derived Quantum from ``a set of experimentally motivated postulates" \cite{cr97} or ``more fundamental principles" \cite{fr17}, which in the light of nonlocality and contextuality acquire a much deeper meaning: By postulating ``absence of conditions ensuring causal evolution" and knowledge which is in principle inaccessible to the human experimenter, they were assuming \emph{Principle Q}.

Quantum means endorsing \emph{Principle Q}, which we assume all the time to found rationally interpersonal relationship. Quantum ``weirdness" is helping us to realize how wonderful ordinary life is.


\begin{references}

\bibitem{jb} J. S. Bell, \emph{Speakable and Unspeakable in Quantum Mechanics}, Cambridge Univ. Press: Cambridge, 1987, 2004.

\bibitem{ks67} S. Kochen and E. P. Specker, The problem of hidden variables in quantum mechanics, Journal of Mathematics and Mechanics 17, 59-87 (1967).

\bibitem{jf} J. Fr\"{o}hlich, Demystifying Quantum Mechanics. The ``ETH Approrach", Thursday Colloquia, Paul Scherrer Institut (PSI), March 21, 2019.

\bibitem{fp} C. A. Fuchs and A. Peres, Quantum Theory Needs No 'Interpretation'. \emph{Physics Today} 53, 3, 70-71 (2000); doi: 10.1063/1.883004


\bibitem{mb26} M. Born, Zur Quantenmechanik der Stoßvorg\"{a}nge. [Vorl\"{a}ufige Mitteilung.$^{1)}$] (Eingegangen am 25. Juni 1926). Zeitschrift f\"{u}r Physik, December 1926, Volume 37, Issue 12, pp 863–867.

\bibitem{bv} G. Bacciagaluppi, A. Valentini, Quantum Theory at the Crossroads: Reconsidering the 1927 Solvay Conference. Part III: The proceedings of the 1927 Solvay conference. Cambridge University Press, 2009. arXiv:quant-ph/0609184v2 (2009).

\bibitem{Guerreiro12} T. Guerreiro, B. Sanguinetti, H. Zbinden, N. Gisin, A. Suarez, Single-photon space-like antibunching, \emph{Phys. Lett. A} \textbf{376}, 2174-2177 (2012); A. Suarez, Empty waves, many worlds, parallel lives, and nonlocal decision at detection, arXiv:1204.1732 (2012).

\bibitem{epr} A. Einstein, B. Podolsky, and N. Rosen, Can Quantum-Mechanical Description of Physical Reality be Considered Complete? \emph{Physical Review}, 47, 777-780 (1935).

\bibitem{hasp} Nicholas Harrigan, Robert W. Spekkens, Einstein, incompleteness, and the epistemic view of quantum states. \emph{Found. Phys.} \textbf{40}, 125 (2010). arXiv:0706.2661 (2007).

\bibitem{db} D. Bohm, A suggested interpretation of the quantum theory in terms of ``hidden'' variables. I and II. \emph{Phys. Rev.} \textbf{85} 166-193 (1952).

\bibitem{szgs} A.Stefanov, H. Zbinden, N. Gisin, and A. Suarez, {\em Phys. Rev. Lett.} 88.120404 (2002); e-print {\em quant-ph/0110117}.
Full length article \emph{Phys. Rev. A} \textbf{67}, 042115(2003).

\bibitem{ng12} N. Gisin, Are There Quantum Effects Coming from Outside Space-time? Nonlocality, free will and ``no many-worlds". In: A. Suarez and P. Adams (Eds.) Is Science compatible with free will? Exploring free will and consciousness in light of quantum physics and neuroscience. Springer, New York, 2013, Chapter 3, pp. 23-39. arXiv:1011.3440v1.

\bibitem{as10a} A. Suarez, Deriving Bell's nonlocality from nonlocality at detection. arXiv:1009.0698(2010).

\bibitem{he} H. Everett, ``Relative" state formulation of quantum mechanics. Rev. Mod. Phys.,29:454-462, 1957.

\bibitem{lev} L. Vaidman, The Many-Worlds Interpretation of Quantum Mechanics. In: \emph{The Stanford Encyclopedia of Philosophy} (Summer 2002 Edition), E. N. Zalta (ed.)

\bibitem{deutsch} D. Deutsch, Lectures on Quantum computation, Lecture2: Interference; Lecture 4: The Schr\"{o}dinger picture, Video 34:48. Cited at August 15, 2017.
 http://www.quiprocone.org/Protected/DD\_lectures.htm

\bibitem{Brassard} G. Brassard and P. Raymond-Robichaud, Can Free Will Emerge from Determinism in Quantum Theory? In: A. Suarez and P. Adams (Eds.) \emph{Is Science compatible with free will? Exploring free will and consciousness in light of quantum physics and neuroscience.} Springer, New York, 2013, Chapter 4, pp. 41-61;

\bibitem{BRR1} G. Brassard and P. Raymond-Robichaud, Parallel lives: A local-realistic interpretation of "nonlocal" boxes, 	 arXiv:1709.10016 (2017).

\bibitem{fr} D. Frauchiger and R. Renner, Quantum theory cannot consistently describe the use of itself, arXiv:1604.07422v2 (2018).

\bibitem{BRR2} G. Brassard and P. Raymond-Robichaud, The equivalence of local-realistic and no-signalling theories,arXiv:1710.01380 (2017).

\bibitem{ng17} N. Gisin, Collapse: what else? arXiv:1701.08300v2 (2017).

\bibitem{Specker60} E. Specker, Die Logik nicht gleichzeitig entscheidbarer Aussagen, Dialectica 14 (1960) 239-246. English translation by M.P. Seevinck: The logic of non-simultaneously decidable propositions, arXiv:1103.4537v3 [physics.hist-ph].

\bibitem{as17} A. Suarez, All-Possible-Worlds: Unifying Many-Worlds and Copenhagen, in the Light of Quantum Contextuality, https://arxiv.org/abs/1712.06448v2 (2017-2019).

\bibitem{aa} A. Albrecht, A cosmos in the lab, \emph{Nature} 542 (2017) 164.

\bibitem{sa17} Scott Aaronson, Is ``information is physical" contentful?, post July 20th, 2017 at 8:30 pm, cited March 16, 2019, https://www.scottaaronson.com/blog/?p=3327

\bibitem{sk15} S. Kochen, A Reconstruction of quantum Mechanics, arXiv:1306.3951v3 (2015).

\bibitem{sk17} S. Kochen, Born's Rule, EPR, and the Free Will Theorem, arXiv:1710.00868 (2017).

\bibitem{vs} V. Scarani, The universe would not be perfect without randomness: a quantum physicist's reading of Aquinas. In:  R. Bertlmann, A. Zeilinger (eds), \emph{Quantum [Un]speakables II}, Springer, 2017, pp. 167-174. https://arxiv.org/abs/1501.00769

\bibitem{ck} J.H. Conway and S. Kochen, The Free Will Theorem. \emph{Found. Phys.} \textbf{36}, 1441-1473 (2006); quant-ph/0604079. The Strong Free Will Theorem. \emph{Notices of the American Mathematical Society} \textbf{56}, 226-232 (2009).

\bibitem{ng15} N. Gisin, Time Really Passes. Science can't deny that. Presentation at ``Workshop on Time in Physics", ETH Z\"{u}rich, September 7-11, 2015.

\bibitem{caf10} C. A. Fuchs, QBism, the Perimeter of Quantum Bayesianism, https://arxiv.org/abs/1003.5209

\bibitem{ma} M. Araujo, quoted in: Ph. Ball, Mysterious Quantum Rule Reconstructed From Scratch, Quanta Magazine, February 13, 2019. Cited at April 18, 2019:
    https://www.quantamagazine.org/the-born-rule-has-been-derived-from-simple-physical-principles-20190213/; Blog, post:
    http://mateusaraujo.info/2019/02/14/the-born-rule-is-obvious/

\bibitem{caf} C. A. Fuchs, On Participatory Realism. arXiv:1601.04360 v3 (2016); What a QBist Painting of Quantum Theory Might Look Like, Contextuality workshop, ETHZ, 22-23 June 2017.

\bibitem{caf17} C. A. Fuchs, Notwithstanding Bohr, the Reasons for QBism, arXiv:1705.03483v1 (2017)


\bibitem{fr17} D. Frauchiger and R. Renner, A non-probabilistic substitute for the Born rule, arXiv:1710.05033 (2017).

\bibitem{mgm} L. Masanes, T.D. Galley, and M.P. Müller, The measurement postulates of quantum mechanics are operationally redundant, \emph{Nature Communications}, volume 10, Article number: 1361 (2019), arXiv:1811.11060 (2019).

\bibitem{cr97} C. Rovelli, Relational Quantum Mechanics, DOI:10.1007/BF02302261, arXiv:quant-ph/9609002v2 (1997).



































\end{references}
\end{document}